# Supervisory Switching Control of an Unmanned Surface Vehicle

Ivan R. Bertaska and Karl D. von Ellenrieder

Department of Ocean & Mechanical Engineering
Florida Atlantic University
Dania Beach, FL, USA
ibertask@fau.edu

*Abstract*—**A novel method to determine the switching of controllers to increase the performance of a system is presented. Three controllers are utilized to capture three "behaviors" representative of unmanned surface vehicles (USVs). An underactuated nonlinear controller is derived to transit the vehicle between locations; a fully-actuated nonlinear controller is given to station-keep the vehicle at a setpoint; and a linear anti-windup controller is presented for the reversing mode of operation. Given a trajectory to follow, a performance-based supervisory switching control system (PBSSC) dictates the switching between controllers to improve system performance. Experimental results are shown that indicate that the PBSSC system is able to mitigate errors in pose better than any of the individual controllers.**

*Keywords—unmanned surface vehicles; supervisory switching control; nonlinear control; control allocation*

## I. INTRODUCTION

For the continued expansion of the roles unmanned surface vehicles (USVs) play in day-to-day operations, multiple in-field behaviors must be taken into account by the vehicles' control architectures. Control systems for USVs often focus on a single control law with certain performance requirements dictating the type, gains, and parameters for the closed-loop system. Although this is an effective way of designing a controller for a particular task, it does not translate well to the responsibilities that unmanned systems will encounter in the near future [1]. USVs will need to follow a wide set operating guidelines as their functions expand in maritime operations (e.g. docking, transiting, berthing, etc.). Often times, the framework for deriving the control laws for different parts of a mission will change as well [2]. This may prompt the need for separate control laws when operational conditions vary across mission objectives, and a means of suitably switching between them, while still preserving stability during transitions. Rather than explicitly commanding these transitions from a high-level motion planner, advantages are presented in redistributing this decision to the low-level control architecture. Namely, combining trajectory segments through control protocol synthesis [3] and computationally expensive control space searches can be avoided, while the ability for immediate control law switching is retained through process or performance estimation.

This work seeks to address the question, given a time-parameterized set of poses $\boldsymbol{\eta}_d(t) = [x_d(t), y_d(t), \psi_d(t)]^T$ incorporating different behaviors a USV may encounter, what combination of control laws provides the best system performance? The control architecture will autonomously select between candidate control laws in real-time judging by their past and immediate performance estimated over a user-defined time window. A framework for interchangeable control laws is proposed through supervisory switching control (SSC) [4] and Lyapunov falsification [5]. Using this method, underperforming or "falsified" control laws are replaced with potentially higher performing alternatives. Three behaviors are examined – transiting, station-keeping, and reversing, each guided by a separate controller. They are formally defined as:

- *Transiting* - the behavior exemplified by a vehicle navigating from one destination to another. An example would be a container ship navigating between ports, following a given trajectory closely. A nonlinear backstepping controller, as proposed in [6], is used.

- *Station-keeping* – regulating the kinematic states of a vehicle to a desired position and orientation. This is demonstrated in a research vessel collecting acoustic data in a general direction at specific geographic location. A nonlinear backstepping controller, as proposed in [7], is employed for this behavior.

- *Reversing* – sternward motion of a vehicle for short durations. Due to the low dynamic range of this behavior, a Proportional-Integral (PI) controller with an anti-windup extension is proposed for surge velocity control, while a Proportional (P) controller is used to regulate vehicle orientation. A nested loop structure is constructed with a Line-of-Sight (LOS) guidance system providing reference outputs to the inner P/PI control loop [8].

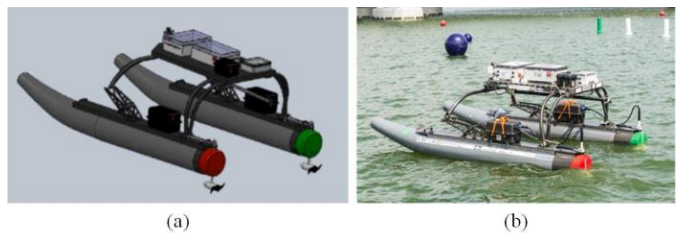

(a)                    (b)

Fig. 1. The USV16 vehicle in transit during the RobotX Maritime Challenge in Marina Bay, Singapore.

The proposed SSC system is validated through on-water experiments on a 4.9m (16') vehicle, the USV16, in the presence of environmental disturbances. The USV16 is a catamaran-hulled vehicle with twin-propellers attached at the transom of each demihull (Fig. 1). These propellers are capable of independently pivoting $\pm 45^o$ with respect to the longitudinal axis of the vehicle, thus making the system overactuated. Due to the actuator configuration, two control allocation schemes are presented for the underactuated and overactuated cases.

This paper is organized as follows: Section II presents background and a brief review on supervisory switching control systems and recent advances in USV control. The vehicle models used for each behavior are found in Section III. Section IV derives the controllers used in these three behaviors. Section V proposes a performance-based supervisory switching control system (PBSSC) capable of selecting controllers *in situ*. Section VI presents results from field experiments. Finally, the paper concludes with some remarks about the future of this topic in Section VII.

## II. BACKGROUND

Supervisory switching control is a form of switching control, and, more generally, is classified as a hybrid system [4]. Hybrid controllers on USVs have been popularized as a means of overcoming Brockett's condition for setpoint stabilization of underactuated systems [9], [10]. It was shown in [11] that underactuated marine vehicles were not stabilizable to an equilibrium with a linear, continuous control law even under body forces, such as gravity. The majority of surface ships in use can be considered underactuated, as there typically is no control authority over the sway degree-of-freedom. Accordingly, controllers have been proposed for the stabilization of an underactuated surface ship at a point, also known as "station-keeping," that feature discontinuous control laws [10], [12], [13], [14]. A subset of these discontinuous control laws focused on the stabilization of these systems under environmental disturbances [15], [16], [17].

Likewise, SSC has also been proposed as a solution to these challenges. SSC employs the use of a "supervisor" to dictate the switching of control laws either through parameter estimation or performance estimation. A controller bank is created through multiple candidate controllers capable of stabilizing the system. These controllers do not need to utilize the same model of the system – only their stability needs to be proven. This allows "off-the-shelf" controllers to be used within the controller bank [4]. SSCs have been proposed as an approach to stabilize marine vehicles with large parametric uncertainty in [18]. Similarly, SSCs were explored as a solution for the varying parameters of remotely operated vehicles in "pick-n-place" operations [19]. Most relevantly in [20], an estimator-based SSC with a human-in-the-loop extension was proposed for multi-objective operations of surface vessels. Although the structure of SSCs for USVs have been explored in theory and simulation, to the authors' knowledge, this is the first work that has published experimental results on a full-sized physical platform.

## III. VEHICLE MODEL

A general model for the three degree-of-freedom (3 DOF) planar motion of a USV is presented in this section. Each controller is derived for a slightly different model of the vehicle, dependent on the operating range of that behavior, and is presented subsequently in Table I.

A 3 DOF system is used to describe the dynamic and kinematic states of the vehicle as in [2]. The typical 6 DOF freedom model is restricted to the surge, sway, and yaw degrees-of-freedom, as the heave, roll, and pitch responses of surface vehicles are negligible in the displacement regime. The combination of rigid-body dynamics and hydrodynamic parameters results in the dynamic model for the remaining states $\boldsymbol{v} = [u, v, r]^T \in \mathbb{R}^3$,

$$\boldsymbol{M}\dot{\boldsymbol{v}} + \boldsymbol{C}(\boldsymbol{v})\boldsymbol{v} + \boldsymbol{D}(\boldsymbol{v})\boldsymbol{v} = \boldsymbol{\tau}, \qquad (1)$$

where the mass parameter matrix $\boldsymbol{M}$ and the centripetal matrix $\boldsymbol{C}(\boldsymbol{v})$ are the summation of rigid-body and added mass parameters, $\boldsymbol{M} = \boldsymbol{M}_{RB} + \boldsymbol{M}_A \in \mathbb{R}^{3\times3}$ and $\boldsymbol{C}(\boldsymbol{v}) = \boldsymbol{C}(\boldsymbol{v})_{RB} + \boldsymbol{C}(\boldsymbol{v})_A \in \mathbb{R}^{3\times3}$. The damping parameter matrix $\boldsymbol{D}(\boldsymbol{v})$ is a combination of the nonlinear drag and linear drag on the system, $\boldsymbol{D}(\boldsymbol{v}) = \boldsymbol{D}_{NL}(\boldsymbol{v}) + \boldsymbol{D}_L$. The control input $\boldsymbol{\tau} = [X, Y, N]^T \in \mathbb{R}^3$ corresponds to the surge force, sway force, and yaw torque applied on the vehicle from the actuators. The pose of the vehicle in the North-East-Down (NED) frame, $\boldsymbol{\eta} = [x, y, \psi]^T \in \mathbb{R}^2 \times \mathbb{S}^1$, is found through the rotation matrix,

$$\begin{bmatrix} \dot{x} \\ \dot{y} \\ \dot{\psi} \end{bmatrix} = \begin{bmatrix} cos\psi & -sin\psi & 0 \\ sin\psi & cos\psi & 0 \\ 0 & 0 & 1 \end{bmatrix} \begin{bmatrix} u \\ v \\ r \end{bmatrix}. \qquad (2)$$

or $\dot{\boldsymbol{\eta}} = \boldsymbol{J}(\psi)\boldsymbol{v}$, transferring the system from the body-fixed coordinate system to the inertial NED coordinate system (Fig. 2).

Values for the hydrodynamic parameters were obtained through a combination of physical modeling and on-water validation [21]. Table I displays the parameter matrices and coefficients in the SNAME 1950 notation [2] for the general model and the models used to derive the three controllers.

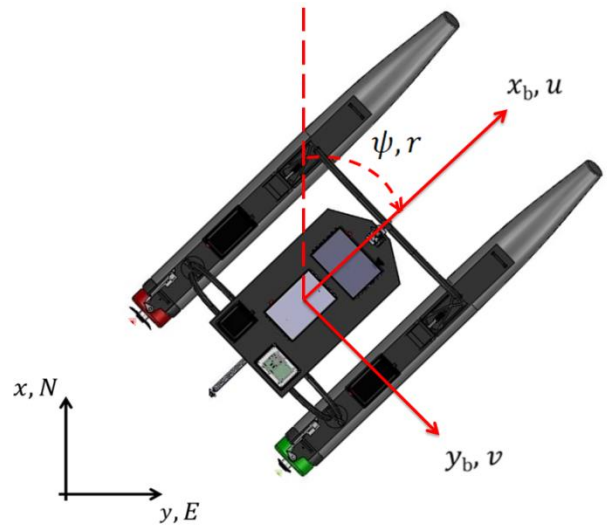

Fig. 2. Top view of the USV16 with both body-fixed (red) and NED inertial (black) coordinate systems.

TABLE I.  MASS PARAMETERS, DRAG PARAMETERS, AND CENTRIPETAL MATRICES FOR GENERAL MODEL AND ALL CONTROLLERS

| | **General Model** | | **Transiting** |
|---|---|---|---|
| $M$ | $M_{gen} = \begin{bmatrix} m - X_{\dot{u}} & 0 & 0 \\ 0 & m - Y_{\dot{v}} & -Y_{\dot{r}} \\ 0 & -N_{\dot{v}} & I_z - N_{\dot{r}} \end{bmatrix}$ | | $M_{tran} = \begin{bmatrix} m - X_{\dot{u}} & 0 & 0 \\ 0 & m - Y_{\dot{v}} & 0 \\ 0 & 0 & I_z - N_{\dot{r}} \end{bmatrix}$ |
| $C(v)$ | $C(v)_{gen} = \begin{bmatrix} 0 & 0 & -(m - Y_{\dot{v}})v + \frac{Y_{\dot{r}} + N_{\dot{v}}}{2}r \\ 0 & 0 & (m - X_{\dot{u}})u \\ (m - Y_{\dot{v}})v - \frac{Y_{\dot{r}} + N_{\dot{v}}}{2}r & -(m - X_{\dot{u}})u & 0 \end{bmatrix}$ | | $C(v)_{tran} = \begin{bmatrix} 0 & 0 & -(m - Y_{\dot{v}})v \\ 0 & 0 & (m - X_{\dot{u}})u \\ 0 & 0 & 0 \end{bmatrix}$ |
| $D(v)$ | $D(v)_{gen} = -\begin{bmatrix} X_{|u|u}|u| + X_u & 0 & 0 \\ 0 & Y_v & Y_r \\ 0 & N_v & N_r \end{bmatrix}$ | | $D(v)_{tran} = -\begin{bmatrix} X_{|u|u}|u| + X_u & 0 & 0 \\ 0 & Y_v & 0 \\ 0 & 0 & N_r \end{bmatrix}$ |
| | **Station-keeping** | | **Reversing** |
| $M$ | $M_{sk} = \begin{bmatrix} m - X_{\dot{u}} & 0 & 0 \\ 0 & m - Y_{\dot{v}} & 0 \\ 0 & 0 & I_z - N_{\dot{r}} \end{bmatrix}$ | | $M_{rev} = \begin{bmatrix} m - X_{\dot{u}} & 0 & 0 \\ 0 & m - Y_{\dot{v}} & 0 \\ 0 & 0 & I_z - N_{\dot{r}} \end{bmatrix}$ |
| $C(v)$ | $C(v)_{sk} = \begin{bmatrix} 0 & 0 & -(m - Y_{\dot{v}})v \\ 0 & 0 & (m - X_{\dot{u}})u \\ (m - Y_{\dot{v}})v & -(m - X_{\dot{u}})u & 0 \end{bmatrix}$ | | $C_{rev} = \begin{bmatrix} 0 & 0 & 0 \\ 0 & 0 & 0 \\ 0 & 0 & 0 \end{bmatrix}$ |
| $D(v)$ | $D_{sk} = -\begin{bmatrix} X_u & 0 & 0 \\ 0 & Y_v & 0 \\ 0 & 0 & N_r \end{bmatrix}$ | | $D_{rev} = -\begin{bmatrix} X_{u,rev} & 0 & 0 \\ 0 & Y_v & 0 \\ 0 & 0 & N_r \end{bmatrix}$ |

## IV. CONTROLLERS AND CONTROL ALLOCATION

Three controllers are described in this section corresponding to the three behaviors proposed for the PBSSC. It is assumed that a dynamically feasible reference trajectory is given to the system as $\eta_d(t) = [x_d(t), y_d(t), \psi_d(t)]^T \in \mathbb{R}^2 \times \mathbb{S}^1$ with a subset $p_d(t) = [x_d(t), y_d(t)]^T \in \mathbb{R}^2$ denoting the desired position in NED.

Both transiting and reversing controllers were developed for an underactuated vehicle with control laws derived separately for the surge thrust $X$ and yaw torque $N$. For a vehicle with differential thrust, as in the USV16, this causes the two laws to compete for control authority from the actuators. To remedy this, an underactuated control allocation system is proposed. The station-keeping control law is developed for a fully-actuated system, and an overactuated control allocation scheme as in [22] is used to transform the control forces and torque into a thrust and azimuth angle for each motor.

### A. Transiting Controller

An underactuated backstepping controller proposed in [6] is used in the transiting case as it has been shown to work well for the planar motion of a vehicle with differential thrust [23]. Aguiar et al. termed this type of controller "position tracking" in that it satisfied a position trajectory $p_d(t)$. Although the closed-loop system could not be deemed to be globally asymptotically stable, it was shown to stabilize the system to an arbitrarily small neighborhood $\delta$.

The dynamic model of the vehicle shown in Eq. (1) with parameters from the transiting case in Table I can be expressed as,

$$\begin{aligned} \dot{p} &= R(\psi)v \\ \dot{v} &= m^{-1}(-S(r)mv - d_v(v)v + gX). \\ \dot{r} &= (I_z - N_{\dot{r}})^{-1}(N_r r + N) \end{aligned} \quad (3)$$

The two dimensional positions $p = [x, y]^T$ and dynamic states $v = [u, v]^T$ can be decoupled from the yaw subsystem since a Nomoto-like, linearized steering model is used for the yaw dynamics [24]. The parameter matrices $m = diag\{m - X_{\dot{u}}, m - Y_{\dot{v}}\}$ and $d_v(v) = diag\{X_{|u|u}|u| + X_u, Y_v\}$ now become submatrices of $M_{tran}$ and $D(v)_{tran}$, respectively. Since this is an underactuated controller, the input parameter matrix $g = [1\ 0]^T$ only influences the surge dynamics. The centripetal matrix $C(v)_{tran}$ is expanded using the skew symmetric matrix,

$$S(r) = \begin{bmatrix} 0 & -r \\ r & 0 \end{bmatrix}, \quad (4)$$

and the two dimensional rotation matrix for coordinate transformation becomes,

$$R(\psi) = \begin{bmatrix} cos\psi & -sin\psi \\ sin\psi & cos\psi \end{bmatrix}. \quad (5)$$

An error vector $p_t$ is expressed in the body-fixed coordinate system denoting the error in position,

$$p_t = R(\psi)^T(p - p_d). \quad (6)$$

Setting an initial Lyapunov function $V_1 = (1/2)p_t^T p_t$ allows for the use of integrator backstepping, as in [25], for control inputs $X$ and $N$. The full control law derivation is excluded here for brevity, but can be found in [6] and [23]. It is important to note that this controller will stabilize a system to a user-defined neighborhood $\delta = [\delta_1, \delta_2]^T$ of the error vector $p_t$.

For clarity, this is expressed as a "ball" matrix in the control law formulation,

$$\boldsymbol{B}(\boldsymbol{\delta}) = \begin{bmatrix} 1 & (m - Y_{\dot{v}})\,\delta_2 \\ 0 & -(m - X_{\dot{u}})\delta_1 \end{bmatrix} \qquad (7)$$

Ultimately, the backstepping procedure produces,

$$X_{tran} = \boldsymbol{g}^T \boldsymbol{\alpha} \qquad (8)$$

where $\boldsymbol{\alpha}$ is a stabilizing function,

$$\boldsymbol{\alpha} = -\boldsymbol{B}(\boldsymbol{\delta})^{-1}[\boldsymbol{h} + \boldsymbol{d}_v(\boldsymbol{v})\boldsymbol{\delta} + m^{-1}\boldsymbol{p}_t + \\ K_{\varphi}m^{-1}\boldsymbol{\varphi}], \qquad (9)$$

$$\boldsymbol{h} = \boldsymbol{d}_v(\boldsymbol{v})\boldsymbol{R}(\psi)^T \ddot{\boldsymbol{p}}_d - K_e \boldsymbol{d}_v(\boldsymbol{v})m^{-1}\boldsymbol{p}_t - \\ m\boldsymbol{R}(\psi)^T \ddot{\boldsymbol{p}}_d + K_e \boldsymbol{z}_1 - K_e^2 m^{-1}\boldsymbol{p}_t. \qquad (10)$$

with positive definite gain matrices $\boldsymbol{K}_{\varphi}$ and $\boldsymbol{K}_e$, and backstepping variables,

$$\boldsymbol{z}_1 = \boldsymbol{v} - \boldsymbol{R}(\psi)^T \dot{\boldsymbol{p}}_d + m^{-1}K_e \boldsymbol{p}_t, \qquad (11)$$

$$\boldsymbol{\varphi} = \boldsymbol{z}_1 - \boldsymbol{\delta}. \qquad (12)$$

Similarly, the yaw control law is found to be,

$$N_{tran} = -\boldsymbol{\varphi}^T m \boldsymbol{B}_b(\boldsymbol{\delta}) - N_r[0\;1]\boldsymbol{\alpha} + [0\;(I_z - \\ N_{\dot{r}})]\dot{\boldsymbol{\alpha}} - K_{z_2} z_2^2, \qquad (13)$$

where $\boldsymbol{B}_b(\boldsymbol{\delta})$ is the second column of Eq. (7), $K_{z_2}$ is a positive gain, and the final backstepping variable is defined as,

$$z_2 = r - [0\;1]\boldsymbol{\alpha}. \qquad (14)$$

Thus, the total control input from the transiting controller is,

$$\boldsymbol{\tau_{tran}} = [X_{tran}, 0, N_{tran}]^T. \qquad (15)$$

Since some parameters in Eq. (13) lead to a control torque greater than the capability of the vehicle, a scaling factor is applied to Eq. (13) to limit the output to a feasible range. Additionally, there will be a trade-off between control effort and performance – the smaller the magnitude of $\boldsymbol{\delta}$, the greater the control effort – which may drive actuators into saturation. These values were carefully tuned in simulation and in on-water field tests.

### B. Station-keeping Controller

A fully actuated MIMO backstepping controller is used as in [26] to command the vehicle during the station-keeping behavior. The control law is derived assuming that all forces are produced at the center of gravity of the vehicle with no actuator constraints. The control allocation scheme in Section IV.E accounts for these limitations to produce an achievable actuator configuration. Like in Section IV.A, the integrator backstepping technique from [25] is used to prove stability and determine the control inputs $\boldsymbol{\tau} = [X, Y, N]^T \in \mathbb{R}^3$. A brief overview of the construction of the control laws is given here. For a full derivation, the authors direct the readers to [7] and [26].

The dynamic model of the vehicle in Eq. (1) with terms from the station-keeping model in Table I is expressed in the inertial coordinate system [2],

$$\boldsymbol{M}_{\eta}(\psi)\ddot{\boldsymbol{\eta}} + \boldsymbol{C}_{\eta}(\boldsymbol{v},\psi)\dot{\boldsymbol{\eta}} + \boldsymbol{D}_{\eta}(\boldsymbol{v},\psi)\dot{\boldsymbol{\eta}} = \boldsymbol{J}(\psi)\boldsymbol{\tau}, \qquad (16)$$

where the inertial parameter matrices are,

$$\begin{aligned} \boldsymbol{M}_{\eta}(\psi) &= \boldsymbol{J}(\psi)\boldsymbol{M}_{sk}\boldsymbol{J}^{-1}(\psi) \\ \boldsymbol{C}_{\eta}(\boldsymbol{v},\psi) &= \boldsymbol{J}(\psi)\widetilde{\boldsymbol{C}}_{sk}\boldsymbol{J}^{-1}(\psi) \\ \boldsymbol{D}_{\eta}(\boldsymbol{v},\psi) &= \boldsymbol{J}(\psi)\boldsymbol{D}_{sk}\boldsymbol{J}^{-1}(\psi) \end{aligned} \qquad (17)$$

and $\widetilde{\boldsymbol{C}}_{sk} = \boldsymbol{C}_{sk}(\boldsymbol{v}) - \boldsymbol{M}_{sk}\boldsymbol{J}^{-1}(\psi)\dot{\boldsymbol{J}}(\psi)$. The error in pose is defined as,

$$\boldsymbol{\eta}_t = \boldsymbol{\eta} - \boldsymbol{\eta}_d. \qquad (18)$$

Two virtual reference trajectories, $\boldsymbol{\eta}_r$ and $\boldsymbol{v}_r$, are now introduced,

$$\dot{\boldsymbol{\eta}}_r = \dot{\boldsymbol{\eta}}_d - \boldsymbol{\Lambda}\boldsymbol{\eta}_t, \qquad (19)$$

$$\boldsymbol{v}_r = \boldsymbol{J}^{-1}(\psi)\dot{\boldsymbol{\eta}}_r. \qquad (20)$$

A diagonal design matrix $\boldsymbol{\Lambda}$ is selected according to desired system robustness and performance [27]. For the controller implemented on the PBSSC, a factor of the estimate of the largest unmodeled delay in the system was used for $\boldsymbol{\Lambda}$ [7]. The reference trajectory was fed into the closed loop system as a series of setpoints, and thus $\ddot{\boldsymbol{\eta}}_d = 0$. This allowed for discontinuities in the reference trajectory without driving the system to instability.

A tracking surface between the vehicle pose and the reference trajectory pose is defined,

$$\boldsymbol{s} = \dot{\boldsymbol{\eta}} - \dot{\boldsymbol{\eta}}_r. \qquad (21)$$

Substituting for $\boldsymbol{s}$ in Eq. (17) produces,

$$\boldsymbol{M}_{\eta}(\psi)\dot{\boldsymbol{s}} = -\boldsymbol{C}_{\eta}(\boldsymbol{v},\psi)\boldsymbol{s} - \boldsymbol{D}_{\eta}(\boldsymbol{v},\psi)\boldsymbol{s} + \boldsymbol{J}(\psi)[\boldsymbol{\tau} - \\ \boldsymbol{M}\dot{\boldsymbol{v}}_r - \boldsymbol{C}(\boldsymbol{v})\boldsymbol{v}_r - \boldsymbol{D}(\boldsymbol{v})\boldsymbol{v}_r]. \qquad (22)$$

Once the equations of motion are in this form, the backstepping procedure as in [26] can be carried out. The resultant MIMO control law is,

$$\boldsymbol{\tau}_{sk} = \boldsymbol{M}_{sk}\boldsymbol{J}(\psi)^T \dot{\boldsymbol{v}}_r + \boldsymbol{C}_{sk}(\boldsymbol{v})\boldsymbol{J}(\psi)^T \dot{\boldsymbol{\eta}}_r + \\ \boldsymbol{D}_{sk}\boldsymbol{J}(\psi)^T \dot{\boldsymbol{\eta}}_r \boldsymbol{J}(\psi)^T K_d \boldsymbol{s} - \boldsymbol{J}(\psi)^T K_p \boldsymbol{\eta}_t, \qquad (23)$$

where $K_p$ and $K_d$ are positive definite gain matrices. Values for the gain matrices were found in simulation and refined in physical experiments on the USV16. The results from these tests are presented in [7].

### C. Reversing Controller

Due to the low dynamic range of the sternward motion of the USV16, a linear controller is able to accurately characterize this behavior. Two linear controllers are developed to control each one of the controllable DOF in the underactuated case – a proportional controller for heading and an anti-windup PI controller for surge velocity. A LOS guidance system is designed to feed reference setpoints to this inner P/PI loop.

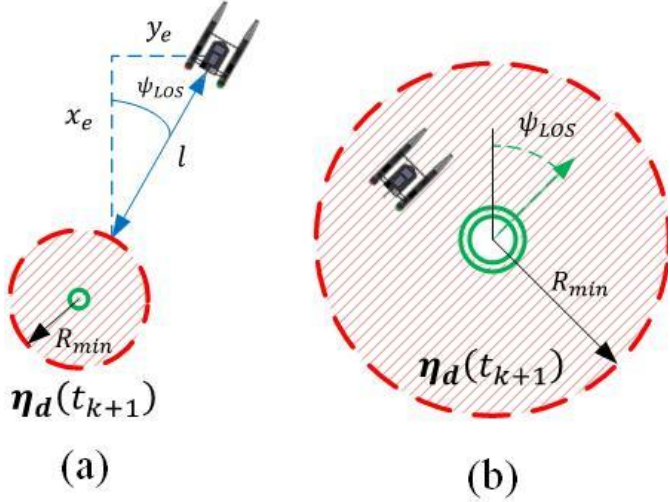

Fig. 3. Calculation to the minimal acceptable distance from the desired waypoint $\boldsymbol{\eta}_d(t_{k+1})$ (red) and corresponding desired heading $\psi_{LOS}$ (blue) (a). When within the minimal acceptable distance, the desired heading $\psi_{LOS}$ matches that of the reference trajectory $\psi_d$ (green) (b). (Figures not to scale)

### 1) Line-of-Sight (LOS) Guidance

A LOS system produces reference outputs $\psi_{LOS}$ and $u_{LOS}$ for the inner P/PI control loop described in the section below. These values are determined through a guidance system similar to the one described in [8].

The desired trajectory $\boldsymbol{\eta}_d(t)$ is discretized into a set of desired poses $\boldsymbol{\eta}_d(t_k) = [x_d(t_k), y_d(t_k), \psi_d(t_k)]^T$ where $k$ is the time index dependent on the update rate of the vehicle. The desired pose used in the LOS system corresponds to the *next* pose in the trajectory $\boldsymbol{\eta}_d(t_{k+1})$, for a time $t \in [t_k, t_{k+1})$. A minimal acceptable distance from the desired pose is defined as $R_{min}$. This is introduced to prevent "looping" behavior, where the vehicle may be forced to double back if the waypoint is overshot. The distance from the vehicle's current position $\boldsymbol{p}(t)$ to the desired waypoint position $\boldsymbol{p}_d(t_{k+1}) = [x_d(t_{k+1}), y_d(t_{k+1})]^T$ can be expressed as,

$$r_t = \left\| \boldsymbol{p}(t) - \boldsymbol{p}_d(t_{k+1}) \right\|_2. \tag{24}$$

Furthermore, the closest radial distance to a circle of radius $R_{min}$ around the desired waypoint (Fig. 3) is,

$$l = r_t - R_{min}. \tag{25}$$

This value will be used to determine both the desired heading $\psi_{LOS}$ and desired speed $u_{LOS}$ for the LOS system.

If the value for $l$ is positive, the vehicle is outside of the minimal acceptable distance from the waypoint, and must be guided towards it. If $l$ is negative, the vehicle is within the minimal acceptable distance and is free to follow the desired heading from the trajectory $\psi_d(t_{k+1})$ (see Fig. 3). The heading reference $\psi_{LOS}$ is determined using this criteria,

$$\psi_{LOS} = \begin{cases} \text{atan2}\left( \frac{y(t) - y_d(t_{k+1})}{x(t) - y_d(t_{k+1})} \right), & l \geq 0 \\ \psi_d(t_{k+1}), & l < 0 \end{cases}, \tag{26}$$

where atan2 refers to the four-quadrant arctan function. Similarly, the desired speed $u_{LOS}$ is defined according to $l$,

$$u_{LOS} = \begin{cases} l/(t - t_{k+1}), & l \geq 0 \\ 0, & l < 0 \end{cases}. \tag{27}$$

Furthermore, instead of explicitly setting the desired velocity $u_{LOS}$ to zero, where the vehicle would exhibit a harsh braking maneuver when entering the minimal acceptable distance, a *kill* command is sent instead. This allowed the low-level control system to set the desired surge thrust $X_{rev}$ to zero, as described in the section below. This LOS system ultimately lead the vehicle to navigate towards the desired waypoint $\boldsymbol{\eta}_d(t_{k+1})$ when out of the minimal acceptable distance, and stop and maneuver to $\psi_d(t_{k+1})$ when within it.

### 2) Heading and Surge Velocity Controller

Since the vehicle model is linearized and the centripetal matrix is cancelled out, the controllable degrees of freedom can be expressed as two SISO transfer functions,

$$\dot{u} = \frac{X_{u,rev}u + X}{(m - X_{\dot{u}})} \Rightarrow \frac{U(s)}{X(s)} = \frac{1}{(m - X_{\dot{u}})s - X_u}, \tag{28}$$

$$\begin{aligned} \dot{\psi} &= r \\ \dot{r} &= \frac{N_r r + N}{(I_z - N_{\dot{r}})} \end{aligned} \Rightarrow \frac{\Psi(s)}{N(s)} = \frac{1}{(I_z - N_{\dot{r}})s^2 + N_r s}. \tag{29}$$

Using Eqs. (28) and (29), one can use the traditional root-locus approach to determine first-cut approximation for gains [28]. The heading subsystem required only proportional control due to the integrator,

$$N_{rev} = -k_\psi(\psi - \psi_{LOS}), \tag{30}$$

for a gain $k_\psi > 0$. The first-order surge subsystem necessitated adding an integration term to achieve the reference setpoint. Thus, a PI controller was selected; however, adding the integration term may saturate the actuators during transient response and result in oscillation about the setpoint. In this case, an anti-windup extension $int_{AW}(u)$ is used to determine the integral value in the control law according to the error in surge velocity, $u_t = u - u_{LOS}$,

$$X_{rev} = \begin{cases} -k_{p,u}(u_t) - k_{i,u} int_{AW}(u, u_{LOS}), & l \geq 0 \\ 0, & l < 0 \end{cases}, \tag{31}$$

for positive gains $k_{p,u} > 0$ and $k_{i,u} > 0$. As was described in the previous section, a null value is set for the surge thrust when the vehicle is within the minimal acceptable distance from the waypoint ($l < 0$). The anti-windup integral function $int_{AW}(u, u_{LOS})$ is only active around a margin of the desired speed $u_{LOS}$ as determined by the factors $0 < \alpha_{min} < 1$ and $\alpha_{max} > 1$,

$$int_{AW}(u, u_{LOS}) = \begin{cases} \sum_{\tau = t'}^{t_k} u_t(\tau) \Delta t, & \alpha_{min}|u_{LOS}| \leq |u| \leq \alpha_{max}|u_{LOS}| \\ 0, & else \end{cases}, \tag{32}$$

where $t'$ is the most recent time the system entered the margin $\alpha_{min}|u_{LOS}| \leq |u| \leq \alpha_{max}|u_{LOS}|$. This results in the integrator resetting when entering the integral-enabled region and being nullified when outside of it, constituting the anti-windup extension to the integral term in the PI controller. The $\Delta t$ refers

to the time step length, or $\Delta t = t_{k+1} - t_k$. Thus, the total control input of the reversing controller is,

$$\boldsymbol{\tau_{rev}} = [X_{rev}, 0, N_{rev}]^T. \tag{33}$$

### D. Underactuated Control Allocation

Since the underactuated controllers in Section IV.A and Section IV.C derive separately the surge and yaw control without considering actuator constraints, these control laws will compete for control authority when implemented on a platform with differential thrust, as in the USV16. This leads to actuator saturation, which results in poor performance or instability. A control allocation system is devised to manage control authority for the transiting and reversing controllers in this section.

A new surge thrust is defined as $X'$ as a function of the steering torque $N$,

$$X' = X e^{-\beta|N|}. \tag{34}$$

The parameter $\beta$ is user-set and dependent on the desired fraction of commanded surge thrust $X'$ during hard corners, where $N = |N_{max}|$, the maximum yaw torque the system is capable of producing. Essentially, this function gives precedence to maintaining a proper heading over speed, leading to a more maneuverable vehicle at the expense of forward motion.

For the underactuated cases, the propellers are positioned so that the thrust is directed parallel with the longitudinal axis of the vehicle $x_b$ (See Fig. 1). The thrust from the port and starboard motors, $T_p$ and $T_s$, is related to the new surge thrust $X'$ and yaw torque $N$ through,

$$\begin{bmatrix} T_p \\ T_s \end{bmatrix} = \begin{bmatrix} 1 & 1 \\ l_{y,p} & -l_{y,s} \end{bmatrix}^{-1} \begin{bmatrix} X' \\ N \end{bmatrix}, \tag{35}$$

where $l_{y,i}, i \in \{p, s\}$ is the lateral distance between the actuator and the vehicle's center of gravity.

### E. Overactuated Control Allocation

Azimuthing thruster configurations such as those found on the USV16 create an overactuated system, since multiple solutions to the controller output $\boldsymbol{\tau}$ can be found in terms of propeller thrust and azimuth angle. This can be formulated as either as a nonlinear optimization problem, where actuator dynamics create constraints on the system, or as a Lagrangian multiplier with extended thrust representation as described in [7] and [22]. Here, the latter approach is applied.

#### 1) Extended Thrust Representation

For the $m$ outputs of the controller, $\boldsymbol{\tau} \in \mathbb{R}^m$, let $\mathbf{f} \in \mathbb{R}^{2r}$ be the actuator forces in the surge and sway directions at each of the $r$ actuators,

$$\mathbf{f} = [F_{x_1} \, F_{y_1} \dots F_{x_i} \, F_{y_i} \dots F_{x_r} \, F_{y_r}]^T. \tag{36}$$

A transformation $\boldsymbol{T} \in \mathbb{R}^{2r \times m}$ from the controller output force $\boldsymbol{\tau}$ to the actuator frame force vector $\mathbf{f}$ can be defined as

$$\boldsymbol{\tau} = \mathbf{T}\mathbf{f}. \tag{37}$$

An extended thrust representation is used to define $\boldsymbol{T}$ [29],

$$\boldsymbol{\tau} = \begin{bmatrix} 1 & 0 & \dots & 1 & 0 \\ 0 & 1 & \dots & 0 & 1 \\ -l_{y_1} & l_{x_1} & \dots & -l_{y_r} & l_{x_r} \end{bmatrix} \begin{bmatrix} F_{x_1} \\ F_{y_1} \\ \vdots \\ F_{x_r} \\ F_{y_r} \end{bmatrix}. \tag{38}$$

The constants $l_{x_i}$ and $l_{y_i}$ represent the longitudinal and lateral distances to the $i$ th actuator measured with respect to the vehicle center of gravity. The solution to the allocation problem now rests in finding an inverse to the rectangular transformation matrix $\boldsymbol{T}$.

#### 2) Lagrangian Multiplier Solution

A cost function $C$ is set up to minimize the force output from each actuator subject to a positive definite weight matrix $\mathbf{W} \in \mathbb{R}^{2r \times 2r}$,

$$\min_{\mathbf{f}} \{ C = \mathbf{f}^T \mathbf{W} \mathbf{f} \}. \tag{39}$$

The optimization problem in Eq. (39) is subject to the constraint $\boldsymbol{\tau} - \mathbf{T}\mathbf{f} = \mathbf{0}$, i.e., the error between the desired control forces and the attainable control forces is minimized. The weight matrix $\mathbf{W}$ is set to skew the control forces towards the most efficient actuators. This is especially important for systems with rudders or control fins, as these actuators provide greater control authority with less power consumption.

A Lagrangrian is then set up as [22],

$$L(\mathbf{f}, \boldsymbol{\lambda}) = \mathbf{f}^T \mathbf{W} \mathbf{f} + \boldsymbol{\lambda}^T (\boldsymbol{\tau} - \mathbf{T}\mathbf{f}). \tag{40}$$

Differentiating Eq. (40) with respect to $\mathbf{f}$, one can show that the solution for $\mathbf{f}$ reduces to $\mathbf{f} = \mathbf{T}_{\mathbf{w}}^\dagger \boldsymbol{\tau}$, where the inverse of the weighted transformation matrix is,

$$\mathbf{T}_{\mathbf{w}}^\dagger = \mathbf{W}^{-1} \mathbf{T}^T (\mathbf{T}\mathbf{W}^{-1}\mathbf{T}^T)^{-1}. \tag{41}$$

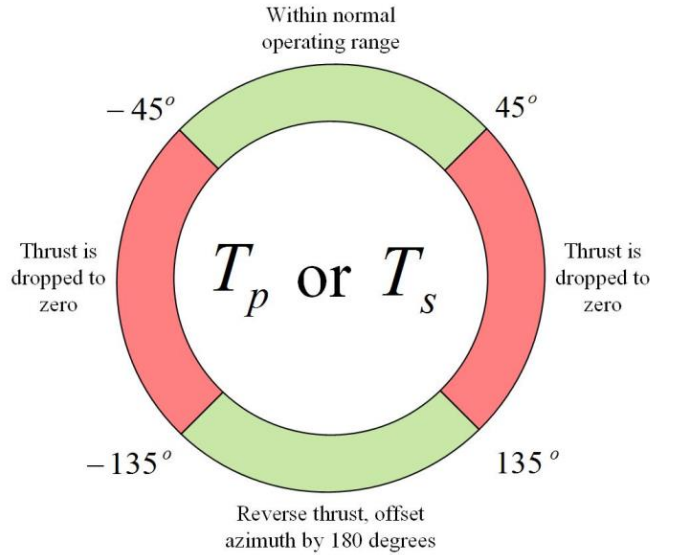

Fig. 4. Motor logic for the control allocation system for the USV16. Achievable configurations are between the $\pm 45^o$ and $\pm 135^o$ azimuths (green), unachievable configurations are outside of that (red), where the motor thrust is dropped to zero to avoid large errors between control output and actuator configuration.

If a vehicle has port/starboard symmetry with identical actuators, the weight matrix $\mathbf{W}$ can be taken as the identity matrix, $\mathbf{W} = \mathbf{I} \in \mathbb{R}^{2r \times 2r}$, and the inverse of the transformation matrix becomes the Moore-Penrose pseudoinverse of the transformation matrix, $\mathbf{T}_\mathbf{w}^\dagger = \mathbf{T}^T(\mathbf{T}\mathbf{T}^T)^{-1}$.

Once the component force vector $\mathbf{f}$ is found, it is trivial to apply a four-quadrant $arctan$ function to find the rotation angle and calculate the magnitude of the thrust at each propeller. Each propeller is capable of achieving a rotation from -45º to 45º – implying that a 180º offset from those values is also attainable by reversing the propeller. A logic scheme is implemented on top of the allocation that stops the thrust if the allocation requests an unachievable angle, and reverses it if an angle from -135º to 135º is given (Fig. 4). This approach produces a computationally efficient answer to the overallocation optimization problem, capable of being implemented on the embedded system within the vehicle [7].

## V. SUPERVISORY SWITCHING CONTROL SYSTEM

A performance-based supervisory switching control system (PBSSC) is proposed to dictate the switching between the transiting, station-keeping, and reversing controllers described in Section IV. The PBSSC uses the direct performance of each controller to choose the most suitable one for the system at the current time. This is accomplished through the falsification of control laws. Controller falsification is built upon the notion that "mental rehearsals" of candidate control laws can be conducted to select the most appropriate one [30]. In [5], Angeli and Mosca introduced a similar concept called "Lyapunov Falsification," which uses a Lyapunov-like function to detect when a controller is approaching instability (i.e., an increase in the Lyapunov function) and select a potentially better performing alternative. The SSC system proposed here uses a combination of both of the above concepts to determine the appropriate controller to improve the performance of the closed-loop system. In this case, simulations of each control law constitute the "mental rehearsals" in [30], and the controller with the smallest accumulated value in a Lyapunov-like function from these simulations is chosen to be inserted in the control loop, similar to [5].

### A. Definitions

A set of candidate controllers $\mathcal{Q} \in \{1, 2, \dots, q_{max}\}$ is defined. Associated with each is a control input $\boldsymbol{\tau}_q \in \mathbb{R}^3$, a pose estimate $\hat{\boldsymbol{\eta}}_q \in \mathbb{R}^2 \times \mathbb{S}^1$, a corresponding closed-loop model $f_q(\boldsymbol{\eta}_d, \boldsymbol{\eta}_q, \boldsymbol{v}_q)$, an error estimate $\hat{\boldsymbol{e}}_q \in \mathbb{R}^2 \times \mathbb{S}^1$, a value from an associated Lyapunov-like function $V_q \in \mathbb{R}$, and a performance signal $\mu_q \in \mathbb{R}$. The inputs to the PBSSC are the desired trajectory, $\boldsymbol{\eta}_d(t) = [x_d(t), y_d(t), \psi_d(t)]^T \in \mathbb{R}^2 \times \mathbb{S}^1$, and the concatenation of the pose and velocity states of the vehicle, represented as the output from the plant, $\boldsymbol{y}(t) = [\boldsymbol{\eta}(t)^T, \boldsymbol{v}(t)^T]^T \in \mathbb{R}^5 \times \mathbb{S}^1$. A switching signal $\sigma \in \mathcal{Q}$ is output, corresponding to the controller selection to be inserted into the closed-loop system.

### B. PBSSC Overview

Fig. 5 displays the basic architecture of the PBSSC. It follows a performance estimator – performance signal generator – switching logic structure. First, in the performance estimator, the performance of each controller over a set time window in the future is represented in a scalar value, $V_q$. This is calculated through simultaneous, real-time simulations of the vehicle with each controller $q$, simulated with its respective closed-loop model, $f_q$. The output of the performance estimator is then fed into a performance signal generator, where the current and previous values of $V_q$ are used to create a new performance signal $\mu_q$. Following the properties of the Lyapunov function, the smallest value of $\mu_q$ would indicate the controller with the best fit. This controller is chosen by the switching logic as the argument of the minimum of the $\mu_q$ set, thereby inserting a best-fit candidate controller into the loop.

### C. The Performance Estimator

The performance estimator uses the output of the system plant $\boldsymbol{y}(t)$ and the desired trajectory $\boldsymbol{\eta}_d(t)$ to simulate the multiple controllers with their corresponding models over a certain time window in the future. This is first accomplished through discretizing the temporal component $t$ into $t_k$, where $k$ refers to the current time step index. A time window $T_K$ is selected to forward simulate the model $K$ time steps into the future. For brevity, the following equations only refer to the current and next time step, but will be repeated similarly until the time step $t_{k+K}$ is reached. The closed loop system output is first estimated using the closed-loop model for each controller,

$$\hat{\boldsymbol{y}}_q(t_{k+1}) = f_q\left(\boldsymbol{\eta}_d(t_k), \boldsymbol{\eta}_q(t_k), \boldsymbol{v}_q(t_k)\right). \tag{42}$$

The estimated error in pose can then be calculated as,

$$\hat{\boldsymbol{e}}_q(t_{k+1}) = \left[\hat{\boldsymbol{\eta}}_q(t_{k+1}) - \boldsymbol{\eta}_d(t_{k+1})\right]^T. \tag{43}$$

A Lyapunov-like function is now defined incorporating this error term,

$$v_q(t_{k+1}) = \tfrac{1}{2}\hat{\boldsymbol{e}}_q^T(t_{k+1})\boldsymbol{P}\hat{\boldsymbol{e}}_q(t_{k+1}). \tag{44}$$

where $\boldsymbol{P} \in \mathbb{R}^{3 \times 3}$ is a positive definite weight matrix. This process is repeated for the next $K$ time steps. The accumulated Lyapunov values over this time window represents the output of the performance estimator and is defined as,

$$V_q(t_{k+1}) = \sum_{\tau = t_{k+1}}^{t_{k+K}} v_q(\tau). \tag{45}$$

### D. Performance Signal Generator

The performance signal generator uses the current and past values of $V_q$ to determine a performance signal $\mu_q$ sent to the switching logic. This allows the past estimated performances of each controller to be factored into the current controller selection. A second time window is now defined as $T_L = t_k - t_{k_0}$, indicating the number of time steps to examine from the past accumulated Lyapunov function values. The performance signal can now be expressed as,

$$\mu_q = \alpha V_q(t_{k+1}) + \beta \sum_{\tau = t_{k_0}}^{t_k} l^{-(\tau - t_{k+1})} V_q(\tau), \tag{46}$$

where $\alpha, \beta > 0$ are tunable parameters weighing the current and past values of the systems, and $l \in [0,1]$ is a "forgetting factor" that allows the user to weigh recent values of $V_q$ more significantly than earlier values.

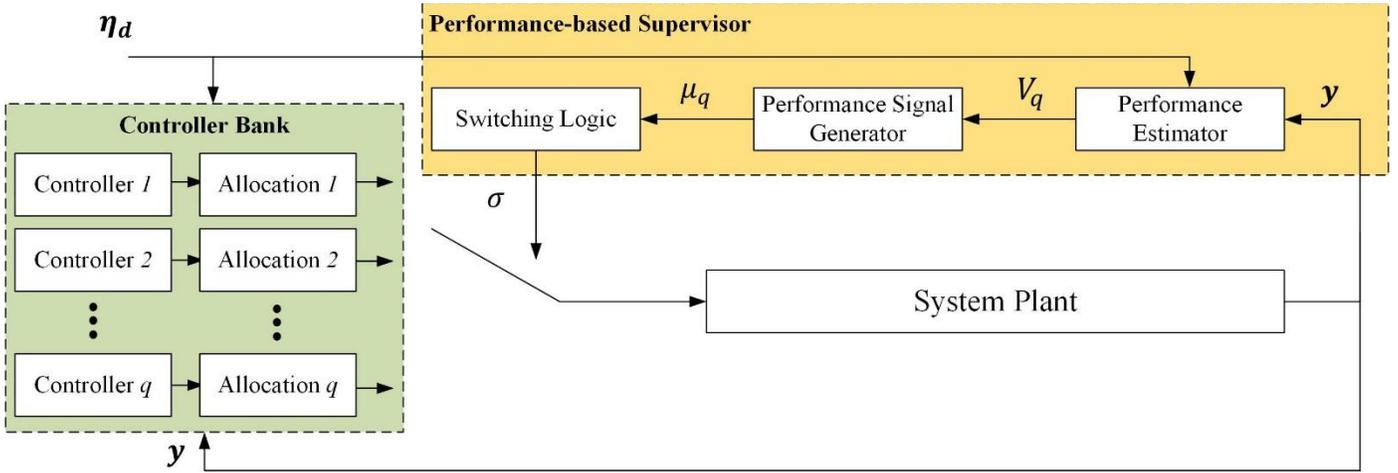

Fig. 5. Performance-based supervisory switching control system for switching amongst $q \in \mathcal{Q}$ controllers. The immediate Lyapunov performance estimate $V_q$ is used to calculate a monitoring signal $\mu_q$ factoring in past performance estimates. The switching logic takes the minimum of the set as $\sigma = \arg \min \mu_q \; \forall q \in \mathcal{Q}$.

### E. Switching Logic

The switching logic determines the controller to be inserted into the loop via the switching signal $\sigma \in \mathcal{Q}$. A temporary value of the switching signal, $\sigma'$, is first calculated as the argument of the minimum of the $\mu_q, q \in \mathcal{Q}$ set,

$$\sigma'(t_{k+1}) = \arg \min_{q \in \mathcal{Q}} \mu_q. \quad (47)$$

Hysteresis is then incorporated into the system to prevent the rapid switching of controllers,

$$\sigma(t_{k+1}) = \begin{cases} \sigma'(t_{k+1}), & (1+h)\mu_{\sigma'(t_k)} \leq \mu_{\sigma(t_k)}, \\ \sigma(t_k), & else \end{cases} \quad (48)$$

for a hysteresis constant $h > 0$. The resultant $\sigma$ is used to insert the corresponding candidate controller into the closed-loop system at the next time step, $t_{k+1}$.

## VI. EXPERIMENTAL RESULTS

### A. Setup

The USV16 was outfitted with a guidance, navigation, and control hardware suite as described in [31] and [32]. Included in this platform was an inertial measurement unit, tilt-compensated compass, as well as a lower level single board computer (SBC), the Technologic Systems TS7800. A higher level SBC was added to provide additional computational capacity, the NVIDIA Jetson TK1. The entire package utilized the Lightweight Data Communication and Marshaling (LCM) system in its underlying architecture [33].

A trajectory was generated to compare the response of the PBSSC with all three controllers against the response of each controller individually. The trajectory was segmented into five parts as shown in Fig. 6a. Circles with arrows denote a "hold and point" segment with the desired heading in the direction of the arrow. Straight arrows without circles depict a move in location with the desired heading also in the direction of the arrow. First, the trajectory required the vehicle to remain in the same location while pointing East for 30s, constituting a station-keeping maneuver (1). It was then tasked with moving due East, slowly accelerating to 1m/s, for 80m (2), immediately followed by holding that position while facing West for another 30s (3). It was then desired to transit back to the starting location at 1m/s (4), subsequently holding the concluding pose for a final 30 seconds (5). It is important to note that this trajectory was feasible for all three controllers to accomplish, in varying degrees.

All experiments were performed at North Lake in Hollywood, FL. This location is a semi-sheltered site off of the US Intracoastal Waterway subject to mild environmental disturbances (wind, current, and waves), which factored into results, but were not debilitating for the vehicle.

### B. Results

For segment (1) of the trajectory, the vehicle used a combination of the transiting, station-keeping, and reversing controller to hold at the initial location for the full 30s (Fig. 6b). Since the vehicle was under some environmental disturbances, it would slowly drift from the desired position, using the reversing controller to back into the location, and the transiting controller to pull into it. The station-keeping controller was employed when there was a lateral error, since this controller enabled the azimuthing of the thrusters. Once at the location, all three controllers were used to maintain the desired heading. After the initial 30s, the vehicle progressed to segment (2), where the PBSSC switched to the transiting controller exclusively, bringing the vehicle up to the nominal 1m/s desired velocity. Once it transited approximately 70m in segment (2), the reversing controller was selected to "brake" the vehicle for the "hold and point" maneuver in segment (3). Since the desired heading in segment (3) was opposite of that in segment (2), the reversing controller was used to align the vehicle in the proper orientation and reverse to the desired point. For segment (3), the vehicle utilized the station-keeping controller to hold that pose, along with the reversing controller when differential thrust was sufficient in orienting the vehicle correctly. The PBSSC once again selected the transiting controller for the return to the initial location in segment (4). The final hold maneuver in segment (5) used a combination of the three controllers to maintain the desired pose, as in segment (1).

Although only one of the field tests is displayed in Fig. 6b, it is representative of all tests run with the PBSSC employed. The same trajectory described in Fig. 6a was run with each controller exclusively to compare the results of the PBSSC with the individual controllers. Two metrics were chosen as the integral of the square of the errors in position and integral of the square of the errors in heading,

$$\Pi_r = \int_0^{t_{final}} \|\boldsymbol{p}(\tau) - \boldsymbol{p}_d(\tau)\|_2^2 \, d\tau, \tag{49}$$

$$\Pi_\psi = \int_0^{t_{final}} \big(\psi(\tau) - \psi_d(\tau)\big)^2 \, d\tau. \tag{50}$$

where $t_{final}$ is the final time in the desired trajectory $\boldsymbol{\eta}_d(t)$. The experiment was repeated four times with the PBSSC enabled, and rerun with each controller exclusively up to four times. The average of these performance metrics across all runs for the PBSSC system enabled, the transiting controller individually, the station-keeping controller individually, and the reversing controller individually are shown in Table II. The PBSSC performed the best with the least amount of errors in position and heading when compared to each controller individually. The position error metrics between the PBSSC and the transiting controller were close, as the controller regulated position very well, but naturally, was unable to regulate orientation as desired. The same can be applied to the reversing controller since neither of these controllers were fully capable of regulating the orientation due to their underactuated nature. When compared against the fully-actuated station-keeping controller, the PBSSC performed better in both position and orientation error regulation. This is due to the limitations in actuator configuration on the USV16. When a thruster is pivoted so that it no longer aligns with the longitudinal axis of the vehicle, less thrust can be devoted towards forward motion, since the motor may reach saturation. This will cause the vehicle to lag behind the desired position for a sufficiently "fast" trajectory. Orientation may be regulated, but the position error will grow. Depending on these errors, the desired actuator configuration from the control allocation described in Section IV.E may be unachievable, causing greater errors in position and heading. This is the case for the results in the station-keeping controller in Table II. Overall, the results demonstrate that the PBSSC-enabled system was better equipped to mitigate errors than any individual controller for this trajectory.

TABLE II. RESULTS OF ON-WATER EXPERIMENTS AVERAGED ACROSS ALL RUNS. TRAN INDICATES TRANSITING, SK STATION-KEEPING AND REV REVERSING

|  | Integral of Square of Position Error $\Pi_r$ (m²)[a] | Integral of Square of Orientation Error $\Pi_\psi$ (deg²)[a] |
|---|---|---|
| PBSSC | 1142.0 | 98663.02 |
| Tran | 1254.5 | 951679.39 |
| SK | 5899.8 | 242036.43 |
| Rev | 51206.6 | 6210014.39 |

[a]. Averaged over all runs.

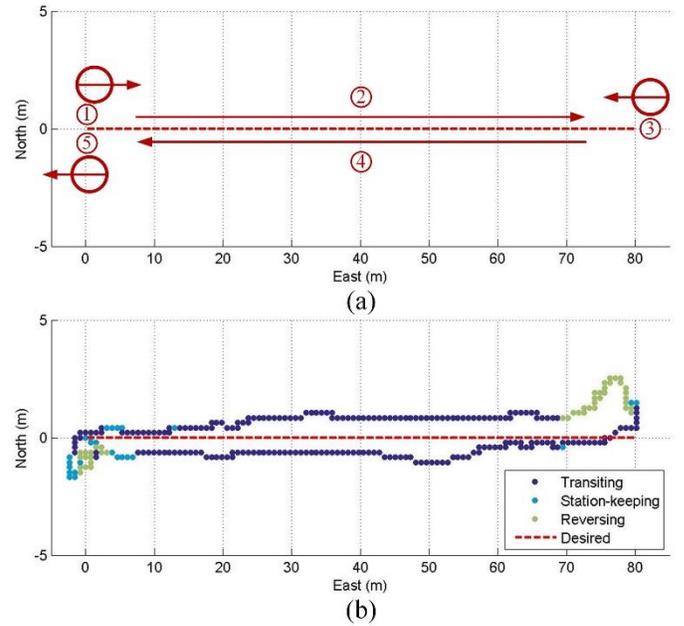

Fig. 6. The desired trajectory (dashed, in red) broken down by segment (a). A representative example of the trajectory of the vehicle with the PBSSC enabled (b). Portions of the trajectory with each individual controller enabled are displayed in their respective colors found in the legend in the bottom right of the figure.

## VII. CONCLUSION

This work has explored the use of a performance-based supervisory switching control system that is capable of autonomously switching between candidate controllers, representative of different "behaviors" a USV might employ in the field. The controllers examined included an underactuated, backstepping transiting controller, a fully-actuated, backstepping station-keeping controller, and a linear P/PI controller with an anti-windup extension for the reversing mode of operation. Real-time, simultaneous simulations and Lyapunov falsification were used to determine which controller to insert into the closed-loop system through the supervisor. The PBSSC was validated on a full-sized USV platform, the USV16, under environmental disturbances. Results show that the PBSSC-enabled system is able to regulate errors in position and orientation better than any individual controller. Future extensions to this work include adding a nonlinear estimator to approximate the varying parameters between the controller models, and incorporating that into the logic behind the controller selection.

### ACKNOWLEDGEMENTS

The authors would like to thank the continued support of the Link Foundation through the Ocean Engineering & Instrumentation Ph.D. Fellowship program. The authors would also like to thank Edoardo Sarda and Ariel Qu for their assistance in conducting field experiments reported here.